\pdfoutput=1
\documentclass[sigconf]{acmart}

\AtBeginDocument{%
  \providecommand\BibTeX{{%
    \normalfont B\kern-0.5em{\scshape i\kern-0.25em b}\kern-0.8em\TeX}}}

\copyrightyear{2021}
\acmYear{2021}
\setcopyright{acmlicensed}\acmConference[AIES '21]{Proceedings of the 2021 AAAI/ACM Conference on AI, Ethics, and Society}{May 19--21, 2021}{Virtual Event, USA}
\acmBooktitle{Proceedings of the 2021 AAAI/ACM Conference on AI, Ethics, and Society (AIES '21), May 19--21, 2021, Virtual Event, USA}
\acmPrice{15.00}
\acmDOI{10.1145/3461702.3462610}
\acmISBN{978-1-4503-8473-5/21/05}

\begin{document}
\fancyhead{}

\title[Designing Disaggregated Evaluations of AI Systems: Choices, Considerations, and Tradeoffs]{Designing Disaggregated Evaluations of AI Systems:\\ Choices, Considerations, and Tradeoffs}

\author{Solon Barocas}
\email{solon@microsoft.com}
\affiliation{
  \institution{Microsoft}
  \city{New York City}
  \state{NY}
  \country{USA}
}

\author{Anhong Guo}
\email{anhong@umich.edu}
\affiliation{
  \institution{University of Michigan}
  \city{Ann Arbor}
  \state{MI}
  \country{USA}
}

\author{Ece Kamar}
\email{eckamar@microsoft.com}
\affiliation{
  \institution{Microsoft}
  \city{Redmond}
  \state{WA}
  \country{USA}
}

\author{Jacquelyn Krones}
\email{jakrones@microsoft.com}
\affiliation{
  \institution{Microsoft}
  \city{Redmond}
  \state{WA}
  \country{USA}
}

\author{Meredith Ringel Morris}
\email{merrie@microsoft.com}
\affiliation{
  \institution{Microsoft}
  \city{Redmond}
  \state{WA}
  \country{USA}
}

\author{Jennifer Wortman Vaughan}
\email{jenn@microsoft.com}
\affiliation{
  \institution{Microsoft}
  \city{New York City}
  \state{NY}
  \country{USA}
}

\author{W. Duncan Wadsworth}
\email{duwads@microsoft.com}
\affiliation{
  \institution{Microsoft}
  \city{Redmond}
  \state{WA}
  \country{USA}
}

\author{Hanna Wallach}
\email{wallach@microsoft.com}
\affiliation{
  \institution{Microsoft}
  \city{New York City}
  \state{NY}
  \country{USA}
}

\renewcommand{\shortauthors}{Barocas et al.}

\begin{abstract}
Disaggregated evaluations of AI systems, in which system performance
is assessed and reported separately for different groups of people,
are conceptually simple. However, their design involves a variety of
choices. Some of these choices influence the results that will be
obtained, and thus the conclusions that can be drawn; others influence
the impacts---both beneficial and harmful---that a disaggregated
evaluation will have on people, including the people whose data is
used to conduct the evaluation.  We argue that a deeper understanding
of these choices will enable researchers and practitioners to design
careful and conclusive disaggregated evaluations. We also argue that
better documentation of these choices, along with the underlying
considerations and tradeoffs that have been made, will help others
when interpreting an evaluation's results and conclusions.\looseness=-1

\end{abstract}





\maketitle

\section{Introduction}

AI systems can perform differently for different groups of people,
often exhibiting especially poor performance for already disadvantaged
groups~\cite[e.g.,][]{nicol2002children,rodger2004field,tatman2017gender,gendershades,bolukbasi2016man,objectdetectionbias,MachineBias,raji2019actionable,SpeechDisparities,OPVM19}. Several
pieces of work have uncovered such performance disparities by
conducting disaggregated evaluations of AI systems, in which system
performance is assessed and reported separately for different groups
of people, such as those based on race and
gender~\cite{mitchell2019model}. Such evaluations can provide a way to
hold development teams and system owners accountable for system
performance, to decide whether to use or keep using a system, or to
identify potential system modifications that would make it acceptable
to use or keep using a system.\looseness=-1

In this paper, we draw attention to the choices that must be made when
designing a disaggregated evaluation. At a high level, these choices
can be thought of as roughly spanning ``why,'' ``who'' (both who will
design and conduct the evaluation and for which groups of people will
system performance will be assessed and reported), ``when,'' ``what,''
``where,'' and ``how.''  Some of these choices influence the results
that will be obtained, and thus the conclusions that can be drawn;
others influence the impacts---both beneficial and harmful---that a
disaggregated evaluation will have on people, including the people
whose data is used to conduct the evaluation.

Using face-based AI systems\footnote{We use the phrase ``face-based AI
  systems'' to refer to face-detection systems, face-characterization
  systems (e.g., gender or age classifiers), face-verification
  systems, and face-identification systems; the latter two are types
  of face-recognition systems.}  as a running example, we argue that a
deeper understanding of these choices will enable researchers and
practitioners to design careful and conclusive disaggregated
evaluations, better enabling themselves and others to understand the
ways in which AI systems perform differently for different groups of
people. To that end, we highlight some of the key considerations that
underlie the choices that must be made when designing a disaggregated
evaluation. We emphasize that these considerations are not
independent; designing a disaggregated evaluation means making
tradeoffs between considerations. These tradeoffs must be clearly
articulated by the evaluation's designers, so that they and others
know how to interpret its results and conclusions.

Although we use face-based AI systems as a running example, the
choices, considerations, and tradeoffs that we discuss are common to
disaggregated evaluations of many other types of AI systems as
well. However, face-based AI systems have, in many ways, become a
bellwether for AI systems in general. Face-based AI systems are used
throughout society in comparatively low-stakes domains like
advertising~\cite{gillespie2019are} and digital media
management~\cite{GooglePhotos}, as well as in high-stakes domains like
education~\cite{Proctorio}, employment~\cite{HireVue},
healthcare~\cite{SepsisWatch},
security~\cite{FaceID,WindowsHello,SmartGate}, and criminal
justice~\cite{FBI-NGI}. Despite their growing prevalence, their use
remains controversial~\cite{S19}. Much of this debate has been spurred
by a number of high-profile disaggregated evaluations, most notably
the Gender Shades study~\cite{gendershades,raji2019actionable}. In
turn, this debate has helped to highlight some key limitations of
disaggregated evaluations and the dangers of focusing narrowly on
performance disparities~\cite{hassein17against,raji2020saving}. For
these reasons, we believe that face-based AI systems make a
particularly compelling running example. That said, our focus on
face-based AI systems should not be viewed as an endorsement of their
use.

Throughout this paper, we intentionally avoid using the word ``audit''
to refer to the process of assessing an AI system for performance
disparities. In other industries, the word ``audit'' refers to an
official examination with mutually agreed-upon actors,
responsibilities, and expectations. Audits typically consider
procedures and documentation, as well as considering system
outputs. Although a disaggregated evaluation could be a component of
an audit of an AI system, it would likely not be the only
component. Developing a clear definition of what it means to audit an
AI system is an extremely important and much-needed research
direction. We are particularly encouraged by a recent paper by
\citet{raji2020closing}, which proposes the ``Scoping, Mapping,
Artifact Collection, Testing, and Reflection'' (SMACTR) framework for
conducting comprehensive internal audits of AI systems via a rigorous,
multi-stage process. Disaggregated evaluations would most likely fit
within the ``Testing'' stage of the SMACTR framework.\looseness=-1

Lastly, we note that performance disparities are just one type of
fairness-related harm. Researchers have highlighted many other types
of fairness-related harms, such as system outputs that stereotype,
demean, or lead to erasure~\cite{barocas2017problem}. Evaluations that
focus on these types of harms are outside the scope of this paper.

In the next section, we provide an overview of disaggregated
evaluations and their role to date in the context of AI systems. We
then describe the choices that must be made when designing a
disaggregated evaluation. We highlight some of the key considerations
that underlie these choices, as well as the tradeoffs between these
considerations. Finally, we conclude with a short
discussion.

\section{Disaggregated Evaluations}
\label{sec:evaluation}

AI systems are typically evaluated by assessing and reporting one or
more aggregate performance metrics---such as accuracy, precision,
recall, word error rate, perplexity, and root-mean-square
error---calculated using an evaluation dataset. For example, the
National Institute of Standards and Technology (NIST) has conducted
vendor tests of commercially available face-recognition systems for
decades, assessing and reporting aggregate performance metrics like
false positive and false negative rates using evaluation datasets that
are comprised of front-facing mugshots, side-view images, webcam
images, and images taken by photo journalists and amateur
photographers~\cite{NIST11,NIST1N}. However, aggregate performance
metrics can obscure poor performance for groups of people that are not
well represented in an evaluation dataset. For example, consider an
evaluation dataset that contains 100 data points, where 90 data points
are associated with group A and 10 data points are associated with
group B. If a system makes correct predictions for the data points
associated with group A and incorrect predictions for the data points
associated with group B, then the aggregate accuracy of the system
will be 90\% when evaluated using the dataset. But this aggregate
accuracy obscures the fact that there is an absolute performance
disparity of 100\% because the accuracy for the data points associated
with group B is zero.

For this reason, researchers and practitioners seeking to uncover
performance disparities exhibited by AI systems often conduct
disaggregated evaluations~\cite{mitchell2019model}. This practice
draws on and parallels similar practices in other industries. For
example, the U.S. Food and Drug Administration mandates that clinical
trial results be assessed and reported separately for groups based on
race, gender, and age. Disaggregated evaluations have proven to be
remarkably effective at uncovering the ways in which AI systems
perform differently for different groups of people. For instance,
\citet{OPVM19} demonstrated that a system used to enroll patients in a
high-risk care management program assigned different risk scores to
Black and white patients with comparable health statuses, leading to a
large disparity in the proportions of Black and white patients
identified for enrollment; \citet{objectdetectionbias} compared the
accuracies of six object-classification systems using images of
household objects from fifty countries, finding that all six systems
had substantially lower accuracies for images from lower-income
countries and households; and \citet{SpeechDisparities} showed that
five commercially available speech-recognition systems had much higher
word error rates for Black people than for white people.

In the context of face-based AI systems, disaggregated evaluations
have been used by researchers and practitioners to assess and report
the performance of face-recognition systems since the early 2000s,
focusing primarily on groups based on environmental factors like pose
and lighting
conditions~\cite[e.g.,][]{Lee2005,Beveridge2009a,Klare2012} and on
groups based on race, gender, and age~\cite[e.g.,][]{BRJ18,
  Krishnapriya2019,CH+19}. In 2019, NIST finally conducted its own
disaggregated evaluation of commercially available face-recognition
systems, focusing on false positive and false negative rates for
groups based on race, sex, age, and country of
birth~\cite{nist2019}. The resulting report provides an overview of
the literature on performance disparities exhibited by face-based AI
systems.\looseness=-1

The most notable such piece of work is the Gender Shades study, which
used a disaggregated evaluation to show that three commercially
available gender classifiers, a type of face-characterization system,
had higher error rates for women with darker skin tones than for women
or for people with darker skin tones overall~\cite{gendershades}. By
focusing on intersectional groups based on skin tone and gender, the
study demonstrated the need to specifically assess and report system
performance for groups based on multiple factors, drawing on and
highlighting the importance of Crenshaw's work on
intersectionality~\cite{Crenshaw}, which showed that the experiences
of Black women differ from the experiences of women or of Black people
overall.\looseness=-1

As well as being widely cited in the research community, Gender Shades
contributed to ongoing changes to the industry around face-based AI
systems. Follow-up work revealed that the study was effective at
getting the companies responsible for the three gender classifiers to
address the performance disparities~\cite{raji2019actionable}; media
coverage of the study led to greater public awareness of the societal
impacts of face-based AI systems~\cite{FaceRecoNYT,JoyNYT}, spurring
calls to action~\cite[e.g.,][]{safeface} and
legislation~\cite{effbills}; and a company responsible for one of the
three gender classifiers announced that it would no longer develop any
face-based AI systems~\cite{IBM}, while another announced that it
would not sell face-recognition systems to police departments in the
U.S.~\cite{MicrosoftFace}.\looseness=-1

\section{Choices, Considerations, and Tradeoffs}

Although disaggregated evaluations are conceptually simple, their
results, conclusions, and impacts depend on a variety of choices. In
this section, we describe these choices---which can be thought of as
roughly spanning ``why,'' ``who,'' ``when,'' ``what,'' ``where,'' and
``how''---and highlight some of the key considerations that underlie
them. We emphasize that these considerations are not independent, and
discuss some of the tradeoffs that must therefore be made when
designing a disaggregated evaluation.

\subsection{What is the goal of the evaluation?}

When designing a disaggregated evaluation, the first choice that must
be made is the goal of the evaluation. There are three considerations
that underlie this choice. First, is the evaluation intended to
demonstrate the existence or absence of performance disparities? Or is
it intended to uncover potential causes of performance disparities?
For example, the ACLU focused on demonstrating the existence of
performance disparities exhibited by Amazon's
Rekognition~\cite{snow18amazon}, while \citet{SpeechDisparities}
attempted to understand \emph{why} speech-recognition systems had much
higher word error rates for Black people than for white people.

Second, will the evaluation focus on actual performance disparities
experienced by a specific set of people who encountered the system in
the past? Or will it focus on potential performance disparities that
may have generally affected people who encountered the system in the
past or that may generally affect people who will encounter the system
in the future? For example, \citet{MachineBias} investigated the
actual risk scores assigned to specific defendants by Northpointe's
COMPAS recidivism-prediction system, while NIST's disaggregated
evaluation of commercially available face-recognition systems focused
on potential performance disparities~\cite{nist2019}. Designing
evaluations that are focused on a specific set of people is usually
easier than designing evaluations that are intended to be
general.\looseness=-1

Finally, will the evaluation be confirmatory or exploratory?
Confirmatory evaluations are intended to provide conclusive evidence
about performance disparities, while exploratory evaluations are
not. By analogy, confirmatory evaluations are like scientific
experiments---that is, they must posit clear hypotheses to be tested
and they must be designed very carefully so as to minimize the risk of
drawing incorrect conclusions. For example, confirmatory evaluations
must account for all factors that can affect system performance,
including demographic factors, environmental factors, and behavioral
factors. As we describe in Section~\ref{sec:additional}, this can be
challenging. However, a failure to do so can yield results that seem
to indicate the existence of meaningful performance disparities, when
in fact these disparities are simply due to spurious
correlations. Confirmatory evaluations are therefore most feasible
when assessing and reporting system performance for a small number of
particularly salient groups in scenarios where there are only a few
additional factors that can affect system performance. In contrast,
exploratory evaluations are not intended to provide conclusive
evidence, so there is much more flexibility in their design. That
said, they can be used to inform the design of subsequent confirmatory
evaluations. Because it is so difficult to design confirmatory
evaluations, most well-known disaggregated evaluations are best
understood as exploratory evaluations.\looseness=-1

\subsection{Who will design and conduct the evaluation?}

A disaggregated evaluation can be designed and conducted by the
development team(s) responsible for the system or by outside parties,
including consultants, researchers, and journalists. When an
evaluation will be designed and conducted by outside parties, this can
be done in collaboration with the development team(s) or it can be
done without their help or knowledge. In some cases, outside parties
can even conduct a disaggregated evaluation when the development
team(s) would prefer that they not.

Many well-known disaggregated evaluations have been designed and
conducted by outside parties. For example, ProPublica, an independent
newsroom, evaluated Northpointe's COMPAS recidivism-prediction
system~\cite{MachineBias}; researchers from Stanford evaluated
speech-recognition systems from Amazon, Apple, Google, IBM, and
Microsoft~\cite{SpeechDisparities}; and the Gender Shades study was
conducted by a researcher from MIT and a researcher from a company
responsible for one of the three gender
classifiers~\cite{gendershades}. Comparatively few well-known
disaggregated evaluations have been performed by development teams,
perhaps because evaluations of this sort are not usually publicly
disclosed, though this situation may change if documentation
approaches like model cards continue to gain
traction~\cite{mitchell2019model}.\looseness=-1

Given that a disaggregated evaluation's results and conclusions can be
quite troubling---and, in some cases, damning for the the development
team(s) responsible for the system---the independence (either
perceived or real) of an evaluation might affect whether its results
and conclusions are seen as trustworthy. As a result, disaggregated
evaluations that are designed and conducted by outside parties may
carry more credibility. That said, outside parties, especially those
that are not working in collaboration with development teams, might
possess fewer resources and may therefore struggle to design and
conduct evaluations that are as comprehensive.

There are also practical limits to what can be achieved by outside
parties. In most cases, outside parties will only be able to engage
with the system as a black box---that is, observe the system's outputs
when presented with different inputs. Moreover, some systems do not
produce publicly accessible outputs, making it impossible for outside
parties to evaluate these systems. In contrast, development teams will
benefit from the fact that they have a deep understanding of their own
systems, including the components that make up their systems and how
these components fit together, the performance metrics that are
already used to evaluate their systems, and their system's intended
use cases and deployment contexts. Development teams are also better
positioned to design and conduct evaluations that are focused on
identifying immediate opportunities for improvement. That said,
outside parties can possess expertise or perspectives that development
teams lack, allowing them to uncover performance disparities that
might otherwise be overlooked.\looseness=-1

\subsection{When will the evaluation be conducted?}

Many well-known disaggregated evaluations have focused on AI systems
that are already commercially available and, in some cases, widely
used---in large part because these evaluations were designed and
conducted by outside parties, without collaboration from the
development teams responsible for the systems. However, disaggregated
evaluations can also take place before system deployment. Indeed,
development teams---potentially in collaboration with outside parties
who have been provided with pre-deployment access---can use
disaggregated evaluations to determine whether their systems are ready
for deployment, reducing the likelihood of harms.\looseness=-1

\subsection{What system or component(s) will be evaluated?}
\label{sec:what}

AI systems often consist of multiple components, such as
machine-learning models, whose inputs and outputs are
interrelated. For example, a speech-recognition system consists of an
acoustic model that models the acoustics of speech and a language
model that models relationships between words. A disaggregated
evaluation can focus on the performance of a system as a whole or on
the performance of one or more of the system's constituent
components. Focusing on the performance of the system as a whole can
make it easier to uncover performance disparities that will lead
directly to harms when the system is deployed. In contrast, focusing
on the performance of one or more components can make it easier to
uncover potential causes of any system-level performance
disparities.\looseness=-1

If a disaggregated evaluation will be designed and conducted by
outside parties without collaboration from the development team(s)
responsible for the system, then it may not be possible to assess the
performance of its constituent components unless they are also
accessible by outside parties. That said, some design choices may
allow component-level performance disparities to be inferred from
system-level performance disparities. For example, by asking Black and
white people to speak the same sequence of words, the researchers who
evaluated five commercially available speech-recognition systems could
be fairly sure that the higher word error rates for Black people were
due to the acoustic models~\cite{SpeechDisparities}.\looseness=-1

Some AI systems depend on the outputs of other systems. A
disaggregated evaluation can yield distorted results if it does not
account for such dependencies. For example, a face cannot be
identified by a face-identification system unless it is first detected
by a face-detection system. Evaluating the performance of a
face-identification system using only images in which faces have
previously been detected will therefore fail to uncover performance
disparities due to face-detection errors. For this reason, NIST
recommends assessing and reporting performance disparities for every
component of a face-recognition system, as well as any other systems
on which it depends~\cite{NIST1N}.\looseness=-1

\subsection{Where will the evaluation occur?}

A disaggregated evaluation can take place ``in the laboratory'' or
``in situ''---that is, in the system's context of use. For example, it
is possible to evaluate a face-verification system that is used to
grant workplace access to employees by presenting the system with a
set of images or by asking employees and non-employees to attempt to
gain access to the workplace using the system. In the former scenario,
the images may not accurately reflect environmental factors like
lighting conditions or behavioral factors like pose. Moreover, if the
system requires an operator to make decisions (e.g., grant or deny
access) based on the system's outputs, then the behavior of the
operator will not be reflected in the
evaluation~\cite{poursabzi2020human}. In contrast, conducting a
disaggregated evaluation in situ allows for the performance of the
entire sociotechnical system, of which the AI system may be just one
component, to be evaluated~\cite{raji2020closing}. Indeed, the
performance of an AI system in isolation may not reveal much about the
ultimate performance of a sociotechnical system that also involves
human discretion and judgment~\cite{selbst2019fairness}. However,
conducting a disaggregated evaluation in situ can be expensive and may
not be possible if the evaluation will be designed and conducted by
outside parties. In addition, some system components may be difficult
to evaluate in situ because they cannot be isolated from the system as
a whole in a meaningful way. However, as we described in
Section~\ref{sec:what}, some design choices may allow component-level
performance disparities to be inferred from system-level performance
disparities.\looseness=-1

\subsection{What are the factors and groups of interest?}
\label{sec:groups}

There are many different groups of people for which AI systems exhibit
poor performance, including groups based on demographic factors,
sociocultural factors, behavioral factors, and morphological
factors. For example, race, gender, age, facial hair, hairstyle,
glasses, facial expression, pose, and skin tone have all been shown to
affect the performance of face-based AI
systems~\cite{NIST11,NIST1N}. Many well-known disaggregated
evaluations have assessed and reported system performance for a small
number of particularly salient groups based on one or two
factors---often groups that are already disadvantaged. This is because
performance disparities involving such groups may compound existing
injustices~\cite{hellman18indirect}. For example, ProPublica focused
on one factor (race) and two groups of people based on that factor
(Black and white defendants) when evaluating Northpointe's COMPAS
recidivism-prediction system~\cite{MachineBias}, while the researchers
who conducted the Gender Shades study focused on two factors (skin
tone and gender) and multiple intersectional groups based on those
factors~\cite{gendershades}.\looseness=-1

The latter example raises an important consideration---namely
whether disaggregated evaluations should focus on social constructs,
such as race and gender, or on observable properties, such as skin
tone and hairstyle. Unlike skin tone and hairstyle, race and gender
are not objective, inherent properties of people; they are categories
constructed by humans that, by social convention, serve as the basis
for social differentiation. These social constructs so deeply
structure people's understanding of others and of themselves that they
are frequently taken for granted. Yet they are historically and
culturally specific, unstable and contested, and often bound up with
unjust social
hierarchies~\cite[e.g.,][]{benthall19racial,hanna20towards}. Even when
focusing on observable properties, some properties may be of
particular interest because they are thought to serve as proxies for
social constructs (e.g., skin tone as a proxy for race). Focusing on
these properties therefore raises some of the same challenges as
focusing on social constructs. In contrast, other observable
properties (e.g., glasses) may be of interest in their own
right.\looseness=-1

Focusing on social constructs (or their proxies) can be advantageous
because social constructs affect people's lives in ways that are both
profound and mundane. For example, some of the groups that are most
disadvantaged within society are groups based on race and
gender. However, this approach raises challenging questions about the
status of social constructs and the implications of using them to
conduct evaluations of AI systems. Determining which social construct
applies to a person can be both practically difficult and ethically
fraught. In many cases, social constructs like gender are simply
ascribed to people by institutions of authority, as is the case with
much official documentation (e.g., government-issued
identification). Relying on such ascriptions assumes that they are
valid and appropriate. In some cases, it may be possible to ask people
about their group membership---an approach that is especially
important when self-reported information may conflict with official
documentation (e.g., when a person does not identify with the gender
ascribed to them at birth). Yet, in other cases, self-reported
information may not be accessible or available at all. In these
scenarios, it may be tempting to infer group membership from
observable properties. However, making decisions about which
observable properties can serve as reliable proxies for social
constructs is an activity that can be viewed as essentializing,
stigmatizing, or alienating, especially if this is done in a way that
suggests that these properties are common to all members---or only
members---of particular groups based on those social constructs. It
also runs the risk of inaccuracies, especially if group membership is
difficult to infer from observable properties. Moreover, inferring
group membership also raises questions about the ethics of imposing
labels on people~\cite{Bogen,andrus2020we}. Lastly, we emphasize that
people may not want a disaggregated evaluation to be designed and
conducted, no matter the potential benefits to uncovering performance
disparities involving the groups to which they belong. In some cases,
this is because the existence of evaluation datasets that contain
information about group membership can be actively
harmful~\cite{scheuerman19how,raji2020saving}. This raises questions
about who gets to decide whether an evaluation will or will not take
place.\looseness=-1

In contrast, focusing on observable properties sidesteps some of the
difficulties presented by social constructs, provided those properties
are not assumed to serve as reliable proxies for social
constructs. However, it can still be challenging to obtain accurate
information about group membership. For example, how short does a
person's hair need to be to be described as ``short'' and what
normative reasons might there be to be interested this factor if not
its relationship to gender? This approach also makes it difficult to
conclude anything about performance disparities involving social
constructs like gender---yet, as we mentioned above, some of the
groups that are most disadvantaged within society are groups based on
social constructs.\looseness=-1

Just as focusing on aggregate performance metrics can obscure poor
performance for groups of people that are not well represented in an
evaluation dataset, focusing on groups based on single factors can
obscure poor performance for people belonging to intersectional
groups. For example, the Gender Shades study demonstrated that three
commercially available gender classifiers had higher error rates for
women with darker skin tones than for women or for people with darker
skin tones overall~\cite{gendershades}. Disaggregated evaluations
should therefore assess and report system performance for people
belonging to both intersectional and non-intersectional
groups.\looseness=-1

Regardless of whether a disaggregated evaluation focuses on social
constructs or on observable properties, it must be possible to create
an evaluation dataset that can support the goal of the evaluation. For
example, if an evaluation will be confirmatory and focused on
potential performance disparities that may generally affect people who
will encounter the system in the future, then the evaluation dataset
must be roughly balanced across the different groups of interest, with
sufficient data about each. In practice, this can be difficult to
achieve, especially if there are many groups of interest (as is often
the case when focusing on intersectional groups). As a result, it is
much easier to design disaggregated evaluations that focus on a small
number of particularly salient groups based on a small number of
factors.\looseness=-1

\subsection{Which additional factors will be accounted for and how will they be accounted for?}
\label{sec:additional}

In practice, there are many factors that can affect system performance
beyond the factors of interest. Because some of these additional
factors may be correlated---even spuriously---with the factors of
interest, a failure to appropriately account for them can make it
difficult to interpret a disaggregated evaluation's results. For
example, although NIST's disaggregated evaluation of face-recognition
systems revealed a higher false negative rate for images of Asian
people than for images of Black or white people, the resulting report
notes that this performance disparity may actually be due to
between-group differences in the time elapsed between each pair of
images~\cite{nist2019}. As another example, a face-recognition system
might perform worse for some genders than for others because it
exhibits poor performance for people with particular hairstyles that
are thought to be meaningfully correlated with gender. From a
normative perspective, attributing these performance disparities to
gender is reasonable---after all, the correlation is not
spurious---though a failure to account for hairstyle will make it more
difficult to uncover the causes of these performance
disparities.\looseness=-1

There are three ways to account for additional factors. The first is
to make sure that the evaluation dataset is reflective of the
population of interest---that is, people who encountered the system in
the past or people who will encounter the system in the future---and
the environmental factors, behavioral factors, and other factors found
in situ. If this is the case, then provided there are no spurious
correlations with the factors of interest---a big assumption---any
performance disparities can be interpreted as being reflective of
those that either affected or will affect the population of
interest. In practice, though, there may well be spurious
correlations, making it difficult to be certain that any performance
disparities are meaningful (or to uncover their potential causes). As
a result, this approach is suitable for exploratory, but not
confirmatory, evaluations. Despite these limitations, many well-known
disaggregated evaluations have used this approach as it is
comparatively easy to
implement~\cite[e.g.,][]{MachineBias,OPVM19}.\looseness=-1

The second way to account for additional factors is to hold their
values constant. For example, when evaluating a face-recognition
system, one way to account for glasses is to only assess and report
performance for people without glasses. Although this approach means
that any performance disparities are more likely to be genuinely due
to the factors of interest, it obscures performance disparities that
occur in the context of factor values other than the ones
considered. Conclusively determining the absence of performance
disparities is therefore especially difficult when using this
approach, making it unsuitable for some confirmatory evaluations. For
example, glasses might have a greater negative effect on system
performance when they are worn by women than when they are worn by
men. A failure to assess and report performance for people wearing
glasses will obscure this performance disparity. As another example,
poor lighting conditions have a greater negative effect on the
performance of face-detection systems for people with darker skin
tones than for people with lighter skin tones. Conducting a
disaggregated evaluation in only good lighting conditions will obscure
this performance disparity. In some cases, it may not even be possible
to hold the values of additional factors constant, as is the case with
evaluations that are focused on actual performance disparities
experienced by a specific set of people who encountered the system in
the past.\looseness=-1

The third way to account for additional factors is to consider a range
of values for each such factor. By assessing and reporting performance
separately for each group of interest and combination of additional
factor values, this approach yields a granular and high-dimensional
view of system performance. Moreover, if the distributions over
additional factor values are the same for each group of interest, then
any performance disparities are more likely to be genuinely due to the
factors of interest. In practice, though, it may be difficult to
ensure that the distributions over additional factor values are the
same for each group of interest. For example, some hairstyles are more
common among some genders than among others. However, even if the
distributions over additional factor values are not the same, it is
still possible to account for them using statistical techniques like
generalized linear mixed-effect regression
models~\cite[e.g.,][]{Givens2013,Beveridge2010}, as we describe in
Section~\ref{sec:performance_analysis}. We note that it can be
difficult to identify all relevant additional factors or to choose an
appropriate range of values for each such factor.  That said, this
approach will most conclusively demonstrate the existence or absence
of performance disparities, making it particularly appropriate for
confirmatory evaluations.\looseness=-1

Lastly, we emphasize that no group is a monolith, so there will always
be additional factors that exhibit within-group variation. For
example, people belonging to any group based on race will exhibit a
wide range of skin tones, not to mention genders and ages. If there is
any chance that these factors can affect system performance, then they
should be accounted for by taking the third approach---that is, by
considering a range of values for each such factor.\looseness=-1

\subsection{How will the evaluation dataset be created?}

Because a disaggregated evaluation depends so heavily on the
evaluation dataset used to conduct it, this is one of the most
important choices, with many underlying considerations and
tradeoffs. There are four main approaches to creating an evaluation
dataset. The first is to reuse an existing dataset that was previously
created for some other purpose; the second is to create a new dataset
using data scraped from the web or from other data sources; the third
is to create a new dataset using data from the system's context of
use; and the fourth is to create a new dataset by collecting data
directly from data subjects.\looseness=-1

For each approach, we discuss the following considerations and the
tradeoffs between them: time, financial cost, suitability for the
evaluation's goal, representation of the groups of interest, labeling
of group membership, representation of additional factors, labeling of
additional factor values, licensing, consent, and compensation. Some
of these considerations (e.g., time, financial cost, and licensing)
are standard when reusing or creating a dataset, while others (e.g.,
representation of additional factors) influence a disaggregated
evaluation's results and conclusions. Others still (e.g., labeling of
group membership, consent, and compensation) influence the impacts
that a disaggregated evaluation will have on people, including the
people whose data is used to conduct the evaluation.\looseness=-1

\subsubsection{Approach 1: Reuse an existing dataset}
\hfill
\vspace{0.5em}
\break
\noindent
Reusing an existing dataset, such as the IARPA Janus
Benchmark A (IJB-A) dataset for evaluating face-detection and
face-recognition AI systems~\cite{Whitelam2017} or the Adience dataset
for evaluating age and gender classifiers~\cite{Eidinger2014}, is less
time consuming and less expensive than creating a new
dataset. However, existing datasets may not contain sufficient data
about the groups of interest, especially if these groups are
intersectional or rare. For example, the researchers who conducted the
Gender Shades study~\cite{gendershades} analyzed both the IJB-A
dataset and the Adience dataset, finding that both are predominantly
composed of images of people with lighter skin tones and both contain
very few images of women with darker skin tones (7.4\% of the IJB-A
dataset and 4.4\% of the Adience dataset). As a result, the
researchers were unable to use either dataset to conduct their
disaggregated evaluation, spurring them to create their own
dataset---the Pilot Parliaments Benchmark (PPB) dataset. The Gender
Shades study motivated the creation of other general-purpose datasets
that have better representation of some groups based on race, skin
tone, gender, and age~\cite[e.g.,][]{karkkainen2019fairface}. However,
other groups, especially groups that are intersectional or rare, are
still underrepresented in these datasets.\looseness=-1

Even if existing datasets do have sufficient data about the groups of
interest, the data points may not be labeled so as to indicate group
membership. Combined with the fact that existing datasets rarely
provide mechanisms for contacting data subjects, this means that
reusing an existing data set may require group membership to be
inferred, perhaps by asking crowdworkers or by using the outputs of
other AI systems. However, as we mentioned in
Section~\ref{sec:groups}, inferring group membership can be
problematic due to the possibility of inaccuracies and due to
questions about the ethics of imposing labels on
people~\cite{Bogen,andrus2020we}.\looseness=-1

Existing datasets may also lack appropriate representation of
additional factors. There are several ways in which this might be the
case, depending on the approach used to account for additional
factors. First, the dataset may not be reflective of the population of
interest and the environmental factors, behavioral factors, and other
factors found in situ. If this is the case, then it is not possible to
implicitly account for additional factors via the argument that any
performance disparities are reflective of those that either affected
or will affect the population of interest. For example, when used by a
retailer to count in-store customers, a face-detection system might
rely on images taken by a wall-mounted camera. Evaluating such a
system using a dataset that is composed of well-lit, front-facing
images will likely yield results that are not reflective of the
system's performance in practice. Second, the data points may have
different values for the additional factors, making it impossible to
account for them by holding their values constant. Even if the data
points do have the same value for each additional factor, this value
may not reflect the values typically found in situ. Third, if
additional factors are to be accounted for by considering a range of
values for each such factor, the data points may not span an
appropriate range for each factor or there may not be sufficient data
for some combinations of factor values. In all three cases, the data
points may not be labeled so as to indicate the values of the
additional factors. This is especially likely for additional factors
that are not typically considered ``interesting'' in and of
themselves. Taken together, these limitations mean that most existing
datasets are not appropriate for conducting confirmatory
evaluations.\looseness=-1

In some cases, licensing restrictions may prevent the reuse of
existing datasets, especially for development teams in companies who
wish to understand the performance of their own AI systems. And, given
that existing datasets rarely provide mechanisms for contacting data
subjects, reusing an existing dataset makes it difficult to request
data subjects' consent or to compensate them fairly for contributing
their data~\cite{AG+18}.\looseness=-1

Aside from these considerations, reusing an existing dataset can also
lead to problems if the dataset is not well documented, for example,
with a datasheet~\cite{gebru2018datasheets}. This is because the
dataset may have been collected or preprocessed in ways that might
affect an evaluation's results but are not immediately apparent.

\subsubsection{Approach 2: Create a new dataset using scraped data}
\hfill
\vspace{0.5em}
\break
\noindent When an appropriate existing dataset does not exist, an
alternative is to create a new dataset using data scraped from the web
or from other sources.  This approach is more time consuming than
reusing an existing dataset, and can be more or less
expensive. Provided the data sources are selected carefully, scraping
data can make it easier to ensure sufficient data about the groups of
interest. For example, when creating the PPB dataset, the researchers
who conducted the Gender Shades study opted to scrape images from the
parliamentary websites of six different countries, which were chosen
because their parliaments were roughly balanced in terms of gender. To
ensure sufficient data for both darker and lighter skin tones, three
of the countries (Rwanda, Senegal, and South Africa) were in Africa,
while the other three (Iceland, Finland, and Sweden) were in
Europe. This process resulted in a dataset that is roughly balanced
across intersectional groups based on skin tone and gender. We note
that creating a new dataset using data scraped from the web or from
other data sources often requires labeling the data points so as to
indicate group membership, again raising the possibility of
inaccuracies, as well as questions about the ethics of imposing labels
on people.\looseness=-1

Scraping data can sometimes make it easier to ensure appropriate
representation of additional factors, though most datasets created
using scraped data will still not be capable of supporting
confirmatory evaluations. For example, by scraping images from
parliamentary websites, the researchers who conducted the Gender
Shades study also ensured that the values of some additional factors
(i.e., lighting conditions and pose) were held reasonably constant;
the images were not, however, labeled so as to indicate the values of
these factors. Other additional factors, such as age, facial
expression, facial hair, hairstyle, glasses, and image resolution,
were not accounted for in any way.\looseness=-1

Using data scraped from the web or from other
data sources also raises a number of considerations involving licensing,
consent, and compensation~\cite{Exposing.ai}. First, it can be difficult to determine
whether a data source is licensed in a such a way that data can be
legally scraped from it and used to conduct a disaggregated
evaluation. Second, even if a data source is appropriately licensed,
this does not mean that the data subjects have consented to their data
being used in this way~\cite{raji2020saving}. For example, in the context of face-based AI
systems, image licenses are typically chosen by the photographer, not
the people depicted in them. A failure to obtain consent from data
subjects may even result in legal action in some jurisdictions. For
instance, the Illinois Biometrics Information Privacy Act has
generated significant litigation over whether and how companies may
collect or use Illinois residents' biometric
information~\cite{law360}. Lastly, even if a data source is
appropriately licensed and all data subjects have consented to their
data being used to conduct a disaggregated evaluation, most data
sources do not provide mechanisms for contacting data subjects, making
it very difficult to compensate them fairly for contributing their
data~\cite{AG+18}.\looseness=-1

\subsubsection{Approach 3: Create a new dataset using data from the system's context of use}
\hfill
\vspace{0.5em}
\break
\noindent
The third approach is to create a new dataset using data from the
system's context of use. If the system has already been deployed, this
can mean reusing deployment data from the system itself. For example,
when evaluating a high-risk care management enrollment system,
\citet{OPVM19} used data about patients that had already interacted
with the system. If, however, the evaluation will take place before
system deployment, then the data must be obtained via other
methods. For example, when evaluating child welfare risk models prior
to their potential deployment in Allegheny County, \citet{CP+18} used
records that were previously collected by Allegheny County to assess
and report different models' area under the curve for groups based on
race.\looseness=-1

Creating a new dataset using data from the system's context of use can
be more time consuming than reusing an existing dataset, and more or
less time consuming than creating a new dataset using scraped data.
The financial cost can vary. We note that if an evaluation will be
focused on actual performance disparities experienced by a specific
set of people who encountered the system in the past, then using data
from the system's context of use is the only option.\looseness=-1

Data from the system's context of use will be reflective of the
population of interest and the environmental factors, behavioral
factors, and other factors found in situ, provided there are no
substantial changes over time. However, there may not be sufficient
data about some groups of interest and the data points may not be
labeled so as to indicate group membership. In some cases, it may be
possible to contact data subjects.

Because data from the system's context of use will be reflective of
the population of interest and the environmental factors, behavioral
factors, and other factors found in situ, it is possible to implicitly
account for additional factors via the argument that any performance
disparities are reflective of those that either affected or will
affect the population of interest. However, accounting for additional
factors by holding their values constant or by considering a range of
values for each such factor is likely not possible, making it
difficult to conduct confirmatory evaluations using such datasets. As
with existing datasets and new datasets created using scraped data,
the data points may not be labeled so as to indicate the values of the
additional factors, although they may be in some cases.\looseness=-1

Data from the system's context of use is unlikely to be subject to
licensing restrictions that prevent it from being used by the system's
development team(s). However, there may be licensing restrictions that
prevent it from being used by outside parties. Using data from the
system's context of use also raises questions about consent and
compensation. Although it may be possible (though not necessarily
easy) to contact data subjects to request their consent or to
compensate them fairly for contributing their data, the perceived need
to do this may be weaker than when creating a new dataset by
collecting data directly from data subjects.\looseness=-1

\subsubsection{Approach 4: Create a new dataset by collecting data}
\hfill
\vspace{0.5em}
\break
\noindent
By far the most flexible approach is to create a new dataset by
collecting data directly from data subjects. However, this approach is
much more time consuming and much more expensive than the approaches
described above. Indeed, the financial costs can be so great that this
approach may only be feasible for development teams with extensive
resources.\looseness=-1

New datasets created by collecting data directly from data subjects
can support evaluations that will take place in the laboratory or in
situ. One way to achieve the latter is to conduct an in-situ pilot, in
which data subjects interact with a fully or partially functioning
system in tightly controlled conditions. This ensures that the
resulting dataset reflects the system's intended use cases and
deployment contexts, as well as capturing some of the sociotechnical
dynamics and their effects on system
performance~\cite{introna2009facial,raji2020saving}. That said,
in-situ pilots can be time consuming and expensive, limiting
the number of use cases and deployment contexts that can be
considered in practice.\looseness=-1

The flexibility afforded by collecting new data directly from data
subjects means that it is possible to ensure sufficient data about the
groups of interest---even if those groups are intersectional or
rare---and to ask data subjects about their group membership. However,
we note that it is important to make sure that the data collection
mechanisms are sufficiently inclusive of the groups of interest. For
example, \citet{PBKM21} found that many people with disabilities were
unable to capture and upload video or speech samples that could
improve system performance for people with similar disabilities.  It
is also possible to account for additional factors by holding their
values constant or by considering a range of values for each such
factor. And, in either case, the data points can be labeled so as to
indicate the values of the additional factors. New datasets created by
collecting data directly from data subjects are therefore most capable
of supporting confirmatory evaluations. Finally, this approach also
provides complete control over licensing, and makes it easy to request
data subjects' consent and to compensate them fairly for contributing
their data. For example, Facebook AI released a dataset of videos
featuring paid actors who had agreed to participate and had explicitly
provided labels indicating their age and
gender~\cite{facebookdata}. However, as explained by
\citet{raji2020saving}, this approach may place a greater burden on
members of already disadvantaged groups by requiring their
participation as data subjects---itself a form of
injustice.\looseness=-1

\subsection{Which performance metric(s) will be used?}

The performance of many AI systems cannot be usefully summarized via a
single metric. For example, the performance of a face-recognition
system might refer to its false positive rate or its false negative
rate. A disaggregated evaluation that assesses and reports only false
positive rates will fail to uncover disparities involving false
negative rates, and vice versa. Moreover, false positives and false
negatives can cause different harms to different stakeholders,
depending on the use case. If a face-identification system is used to
grant workplace access to employees, then false positives may cause
harms to the employer (e.g., loss of property) by granting access to
non-employees, while false negatives will cause harms to individual
employees (e.g., being unable to do their jobs or feeling singled out
or discriminated against) by denying them access to their workplace.
NIST's disaggregated evaluation of face-recognition systems therefore
assessed and reported false positive and false negative
rates.\looseness=-1

The performance metric(s) that will be used to conduct a disaggregated
evaluation are primarily determined by what will be evaluated (i.e.,
the system as a whole or one or more of its constituent components)
and where the evaluation will take place. For example, when evaluating
commercially available speech-recognition systems,
\citet{SpeechDisparities} used word error rate because they were only
able to access the systems as a whole. Had they been able to access
the systems' language models, they might instead have used
perplexity. As another example, if a face-verification system is
evaluated in situ, then it is possible to use performance metrics that
depend on the operator's decisions.\looseness=-1

We also note that considerations relating to the way in which system
performance will be reported can influence the performance metric(s)
that will be used. For example, a system's false positive rate is one
minus its true negative rate. Therefore assessing a system's false
positive rate is equivalent to assessing its true negative
rate. However, as noted by NIST in a report on its vendor tests of
commercially available face-recognition systems~\cite{NIST11},
``readers don't perceive differences in numbers near 100\% well,
becoming inured to the `high-nineties' effect where numbers close to
100 are perceived indifferently.'' Moreover, the harms experienced by
people whose faces are or are not recognized are directly related to
false positive and false negative rates. In contrast, they are
inversely related to true negative and true positive rates. The report
cites an example that is similar to the workplace access use case
described above and explains how doubling the false negative rate
doubles the number of employees that will be harmed. Reporting the
system's false negative rate, rather than its true positive rate,
therefore emphasizes the direct connection between poor performance
and harms.\looseness=-1

Performance disparities can be reported in absolute or relative
terms~\cite{Keppel2005}. Reporting each group's performance relative
to that of the best-performing group can make it easier to immediately
grasp the extent of any performance disparities. For example, if a
system has a false negative rate of 0.01 for women and a false
negative rate of 0.1 for people who are non-binary, then the absolute
performance disparity is only 0.09, but false negatives are ten times
more likely for people who are non-binary than for
women. \looseness=-1

Lastly, we emphasize that any set of quantitative performance metrics
may fail to capture aspects of system performance that are more
subjective, context-dependent, or otherwise harder to quantify. For
this reason, different people may experience an AI system differently,
even when a disaggregated evaluation appears to demonstrate the
absence of any performance disparities involving the groups to which
they belong.

\subsection{How will performance be analyzed?}
\label{sec:performance_analysis}

The way in which performance will be analyzed depends on many of the
choices that we described above, including the goal of the evaluation,
the way in which additional factors are accounted for, and the
performance metric(s) that will be used.\looseness=-1

The simplest way to analyze performance is to assess each performance
metric for each group of interest and then compare the resulting
values (i.e., point estimates) for each metric. Although this approach
is easy to implement, it does not account for any randomness in the
evaluation dataset and therefore raises the possibility of misleading
results. In contrast, using statistical techniques that incorporate
uncertainty (e.g., $p$-values, confidence intervals, posterior
probabilities, permutation tests) can better mitigate this
possibility. Moreover, some of these techniques make it easier to
detect subtle differences, reducing the likelihood that performance
disparities will go undetected. For confirmatory evaluations, it is
especially important to use techniques that incorporate uncertainty so
as to minimize the risk of drawing incorrect conclusions.\looseness=-1

In the context of face-based AI systems, generalized linear
mixed-effect regression models have been used to conduct both
confirmatory and exploratory
evaluations~\cite[e.g.,][]{Givens2013,Beveridge2010}. As well as
incorporating uncertainty, these models provide a way to account for
additional factors when the distributions over factor values are not
the same for each group of interest. They are also particularly
appropriate when the data points in the evaluation dataset are grouped
(e.g., multiple utterances from the same speaker) and they are well
suited to analyzing performance for intersectional
groups~\cite{Gelman2006b}. However, they often rely on strong
distributional assumptions about the evaluation dataset that may not
hold.\looseness=-1

When conducting an exploratory evaluation, decision trees or other
partitioning methods can be used to construct a granular and
high-dimensional view of system performance. This approach is
particularly effective at uncovering potential causes of performance
disparities that can then be further investigated via subsequent
confirmatory evaluations. For example, researchers have used
high-dimensional performance analyses to understand the effects of
demographic factors, sociocultural factors, and environmental factors
on face-based AI
systems~\cite{Lakkaraju2017,Nushi2018}. High-dimensional performance
analyses can also find intersectional groups for which a system
exhibits poor performance using variable importance ranking or other
variable selection methods. That said, without efforts to mitigate
overfitting, they can yield misleading results.\looseness=-1

It is important to note that all of these approaches will struggle to
distinguish between the effects of factors that are highly
correlated. For example, when assessing and reporting the performance
of a face-recognition system, if all images of people wearing glasses
are poorly lit and all poorly lit images are of people wearing
glasses, but the system only actually exhibits poor performance for
one of these groups, then it will be difficult to conclusively
determine which one. This highlights the benefit of accounting for
additional factors by considering a range of values for each such
factor.\looseness=-1

There are also non-model-based ways to uncover potential causes of
performance disparities. For example, \citet{MP+19} used computer
vision post-processing techniques and post-hoc explanation methods to
uncover potential causes of the performance disparities that were
found in the Gender Shades study.

\subsection{How transparent will the evaluation be?}

Disaggregated evaluations can vary considerably in their level of
transparency. ProPublica made all aspects of their
evaluation of Northpointe's COMPAS recidivism-prediction system
publicly available, including their results and
conclusions~\cite{MachineBias}, a full description of their design
choices~\cite{Larson_Mattu_Kirchner_Angwin_2016}, their evaluation
dataset, and the source code that they used to analyze system
performance~\cite{machinebiascode}. In contrast,
development teams who design and
conduct disaggregated evaluations to understand the performance
of their own AI systems often choose not to disclose any details,
though some have disclosed results and other information as
a way to inform others about their systems' characteristics and
limitations~\cite[e.g.,][]{GoogleModelCard,FaceAPITnote}.

If all aspects of a disaggregated evaluation are made publicly
available, then others can easily repeat the evaluation to verify its
results and conclusions. They can also use the evaluation dataset to
conduct disaggregated evaluations of other AI systems. For example,
many researchers have drawn on ProPublica's design choices, evaluation
dataset, and source code to further interrogate its results and
conclusions~\cite[e.g.,][]{kleinberg2017inherent,corbett2016computer,chouldechova2017fair}. People
may be more likely to trust a disaggregated evaluation if all aspects
are made publicly available, especially if the evaluation's results
are favorable. However, development teams may be reluctant to do this
if making aspects of the evaluation available could provide others
with a competitive advantage.\looseness=-1

Making evaluation datasets publicly available raises the possibility
of dataset misuse, such as using evaluation datasets to develop AI
systems that cause new harms to already disadvantaged groups. The
researchers who conducted the Gender Shades study attempted to
mitigate this possibility by restricting access to the PPB dataset to
researchers who wish to use it for non-commercial purposes. However,
this means that development teams in companies who wish to understand
the performance of their own gender classifiers must recreate the PPB
dataset themselves, and it is not possible do this in a way that
yields an identical dataset.\looseness=-1

\vspace{-0.5em}
\section{Discussion and Conclusion}

We have drawn attention to the variety of choices that must be made
when designing a disaggregated evaluation, as well as some of the key
considerations that underlie these choices and the tradeoffs between
these considerations. Making these choices is rarely an easy task and
their ramifications can be hard to predict. Some of these choices
influence the results that will be obtained, and thus the conclusions
that will be drawn; others influence a disaggregated evaluation's
impacts---both beneficial and harmful---on people, including the
people whose data is used to conduct the evaluation.\looseness=-1

Our paper highlights the importance of taking a careful approach to
designing disaggregated evaluations and can serve as a road map for
evaluation designers. The time and effort spent on dataset
construction, including labeling the values of additional factors that
can affect system performance, will lead to more conclusive---and more
actionable---results. Similarly, ensuring that an evaluation dataset
reflects a system's intended use cases and deployment contexts will
yield results that better reflect the system's performance in
practice. Of course, there are tradeoffs too. For example, tailoring a
disaggregated evaluation too closely to one use case or deployment
context may lead to results that do not generalize.\looseness=-1

Our paper can also serve as a road map for interpreting a
disaggregated evaluation's results and conclusions. Have all factors
that can affect system performance been accounted for? Does the
evaluation dataset reflect the population of interest and the
environmental factors, behavioral factors, and other factors found in
situ?  Was the evaluation confirmatory---that is, did it posit clear
hypotheses to be tested and was it designed very carefully so as to
minimize the risk of drawing incorrect conclusions?  Or is it best
understood as an exploratory evaluation?  However, for our paper to
serve as such a road map, evaluation designers must document their
choices, along with the underlying considerations and the tradeoffs
that they have made. And, as part of that process, they should clearly
communicate any limitations~\cite{raji2020saving}. Existing
documentation approaches, like datasheets for
datasets~\cite{gebru2018datasheets} and model
cards~\cite{mitchell2019model}, may be of value here.\looseness=-1

There are several questions that we have left unaddressed. For
example, we have assumed that an evaluation's designers have access to
``ground truth'' labels for the task that the system is intended to
perform. However, in many scenarios, such labels may be inaccurate,
either due to measurement issues~\cite{jacobs2019measurement} or due
to discrimination~\cite{FSV16}. When this is the case and when there
is no way to identify the extent or nature of the mislabeling, a
disaggregated evaluation's results will be meaningless, no matter how
carefully it was designed. We also emphasize that although uncovering
performance disparities is essential to the responsible deployment of
AI systems, \emph{reducing} performance disparities can be a fraught
task. For example, well-known impossibility results imply that
performance disparities according to one metric may originate from a
decision to reduce performance disparities according to another
metric~\cite{chouldechova2017fair, kleinberg2017inherent}. In
addition, improving performance may not always be a desirable
outcome. This is especially true for face-recognition systems, which
may be perceived as a threat whether they perform well or
poorly~\cite{hassein17against}. Although disaggregated evaluations can
add urgency to normative debates about AI systems, there are many
other considerations beyond performance that can determine a system's
desirability, trustworthiness, or acceptance.\looseness=-1

\section{Acknowledgments}

We thank Natasha Crampton, Todd Glass, Eric Horvitz, Sasa Junuzovic,
Kristen Laird, and Besmira Nushi for their input and feedback.

\balance
\bibliographystyle{ACM-Reference-Format}
\bibliography{references}

\end{document}